# Evaluating the feasibility of using Generative Models to generate Chest X-Ray Data

Muhammad Danyal Malik and Danish Humair
Lahore University of Management Sciences, 25100080, 25100183@lums.edu.pk

*Abstract -* In this paper, we explore the feasibility of using generative models, specifically Progressive Growing GANs (PG-GANs) and Stable Diffusion fine-tuning, to generate synthetic chest X-ray images for medical diagnosis purposes. Due to ethical concerns, obtaining sufficient medical data for machine learning is a challenge, which our approach aims to address by synthesising more data. We utilised the Chest X-ray 14 dataset for our experiments and evaluated the performance of our models through qualitative and quantitative analysis. Our results show that the generated images are visually convincing and can be used to improve the accuracy of classification models. However, further work is needed to address issues such as overfitting and the limited availability of real data for training and testing. The potential of our approach to contribute to more effective medical diagnosis through deep learning is promising, and we believe that continued advancements in image generation technology will lead to even more promising results in the future.

## INTRODUCTION

Chest X-Rays are by far the most common radiographic procedures used for medical diagnosis of lung disorders. Recently, advances have been made at a rapid pace in the field of Computer Aided Diagnosis (CAD), the automated detection of these diseases [1]. However, a notable issue with this approach is the lack of availability of medical data for these purposes. This is largely due to ethical concerns such as the privacy of the patients [2]. This is the problem we aim to solve with our approach. By using image generation models such as GANs and Stable Diffusion, we hope to find a way to utilise existing datasets to produce more data. This can help large classification models such as convolutional neural networks or vision transformers achieve higher accuracy, especially when there is a lack of data. The source code is available on our official GitHub repository (link provided at the end of the paper).

## BACKGROUND

Generative Adversarial Networks (GANs) consist of two neural networks, a generator G and a discriminator D, that are trained simultaneously. The generator takes random noise z as input and generates an image x, while the discriminator evaluates whether x is real or fake. The training objective is to minimise the following loss function:

$$E_x[log(D(x))] + E_z[log(1 - D(G(z)))]$$

Progressive Growing GANs (PGGANs) gradually increase the resolution of the generated images during training. This is achieved by adding new layers to both the generator and discriminator as the training progresses. PGGANs improve training stability and scalability and have shown impressive results in generating high-quality images [3].

Stable Diffusion, on the other hand, is a generative model that learns to generate high-quality images by iteratively refining a noise vector through a series of diffusion steps. Unlike GANs, it does not require adversarial training, making it more stable and less prone to mode collapse [4].

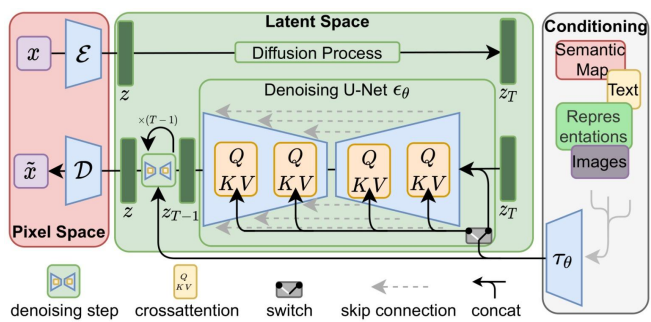

Figure 1
Stable Diffusion Model Overview [5]

Stable Diffusion consists of a diffusion process, which is a sequence of T steps; each step applies a diffusive process that smooths out the noise vector. The smoothing is done using a diffusion process that adds Gaussian noise to the input image. The formula below can be used to obtain the noisy image at a specific time step, *t*.

$$x_t = \sqrt{\alpha_t} x_0 + \sqrt{1 - \alpha_t}\varepsilon$$

The model then generates an image by passing the final smoothed noise vector through a decoder network. The decoder network maps the smoothed noise vector to an image in pixel space.

The training objective of Stable Diffusion is to maximise the log-likelihood of the training data, which is defined as the sum of the negative log-likelihood of the model's predictions over the training data. The model is trained using a maximum likelihood estimation method, which involves minimising the negative log-likelihood of the training data The formula for the reverse diffusion process can be approximated as expressed below.

$$L_{simple} = \mathbb{E}_{t,x0,\varepsilon}\left[\left\|\varepsilon - \varepsilon_\theta(x_t, t)\right\|^2\right]$$

Diffusion has shown impressive results in generating high-quality images and has been applied to various image synthesis tasks, including image inpainting, super-resolution, and image synthesis. It is a promising alternative to GANs and has the potential to generate more stable and diverse images [6].

## METHODOLOGY

### I. Dataset

The dataset we opted to use for these experiments is the Chest X-ray 14 dataset by The National Institutes of Health [7], the primary research facility for conducting medical research in the United States of America. This is a large Chest X-ray dataset available to the public. It comprises the following classes of diseases: No Finding, Atelectasis, Cardiomegaly, Consolidation, Edema, Effusion, Emphysema, Fibrosis, Hernia, Infiltration, Mass, Nodule, Pleural Thickening. Pneumonia and Pneumothorax.

### II. Progressive Growing GAN

PG-GANs (Progressive Growing GANs) were used to synthesise Chest X-rays using the Chest X-ray 14 dataset by Segal et al. [8]. Using their implementation and pre-trained model weights, we generated several Chest X-rays as a baseline for comparison. These included No Finding (no disease) images along with Pneumonia Images. To overcome the issue of not having labels in PG-GANs, the authors used a separate feature-extractor model to generate latent vectors for separate diseases (post-training). Using this approach, they could let the GAN train first and then simply use existing images to generate latent vectors for different diseases.

### III. Stable Diffusion Fine-Tuning

Fine-tuning stable diffusion has been made more accessible than ever since the release of the DreamBooth Notebook [9]. We opted to use the code from this notebook to help us fine-tune Stable Diffusion v2.1-512px on the train set of the Chest X-ray 14 dataset. We intentionally avoided using the test set to avoid exposing our model to the test set, as it would later be used for evaluation. To prepare our data, we needed to add prompts to our images, as stable diffusion consists of both a text encoder and a U-Net. To do this, we needed each image file from the dataset to have a prompt associated with it. In DreamBooth, this can be done by either renaming the files or using text files. In our provided implementation, we have written code to extract the images and each of their labels, rename them based on the diseases found in the image and save them as a new dataset. The implementation and the processed dataset are available on our GitHub Repository. This data was then used to fine-tune the stable diffusion models. The different checkpoints were saved to assess possible overfitting and evaluate the model's performance at each checkpoint during the training process.

### IV. Adding Prompts for Bounding Boxes (Stable Diffusion)

After fine-tuning the model, we had control over what type of disease would be present in the generated image but not where in the image it would be. To add this extra level of control, we utilised the bounding boxes provided in the dataset. These bounding boxes were only present in the test set, so we were not able to utilise as many images. To achieve this, we used a similar method as in sub-section III. Utilising the DreamBooth framework [9] once again, we prepared the data by adding code to extract not only the disease but also a custom label for the position of the finding based on the x and y coordinates of the bounding boxes. These images were once again saved in a folder to use for training, using the same process as highlighted above. This would allow us to generate images where we can explicitly specify the position of the finding, for example, 'top left'.

## RESULTS

### I. Real Images from Dataset

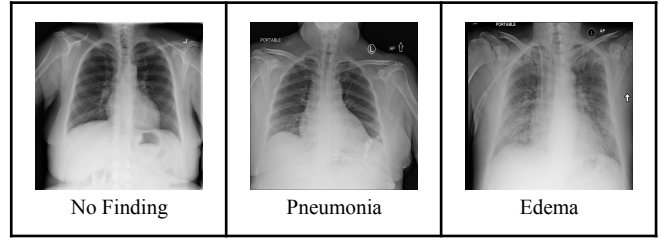

No Finding | Pneumonia | Edema

### II. PG-GAN Images

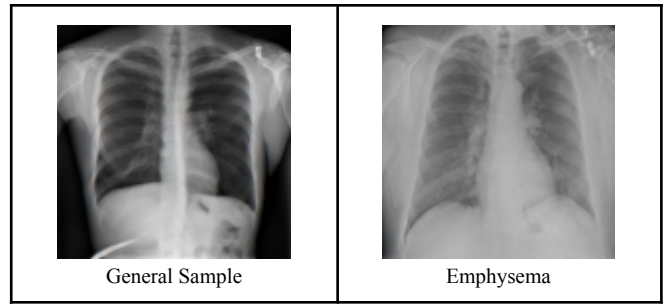

General Sample | Emphysema

*III. Stable Diffusion Images*

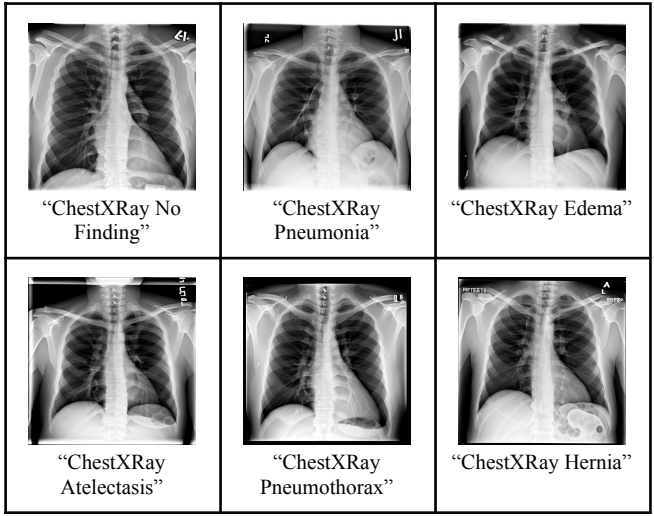

"ChestXRay No Finding" | "ChestXRay Pneumonia" | "ChestXRay Edema"

"ChestXRay Atelectasis" | "ChestXRay Pneumothorax" | "ChestXRay Hernia"

*IV. Stable Diffusion Images with Positions Specified*

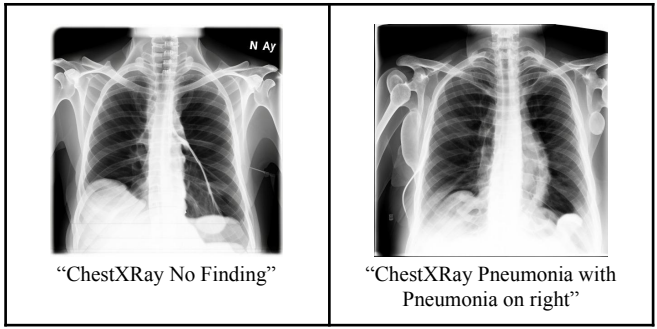

"ChestXRay No Finding" | "ChestXRay Pneumonia with Pneumonia on right"

## EVALUATION

*II. Qualitative Analysis*

From a visual standpoint, all of the images generated look extremely convincing, except for perhaps the position-specific images. This is likely due to the small number of samples available with bounding boxes (<1000) as compared to the data available without bounding boxes (>80,000 in the train set). As for the regular stable diffusion model, the results produced have quite a high resolution as compared to the PG-GAN results. However, they do look less convincing when viewed side-by-side. Overall, both the PG-GAN and the Stable Diffusion results show impressive detail and variation in the samples produced.

*II. Quantitative Analysis*

As an objective analysis of the realism of the generated images, we trained a classification model on a small subset of both real images and real + synthesised images. We kept the overall size of the subset the same for both tests. The model was evaluated on a subset of the test set with a size of 300. This was done for one particular disease, Edema. Hence, the task was to classify the image as either No Disease or Edema. Please note that this test was only performed for our original fine-tuned stable diffusion model, i.e. Methodology section, sub-section III. The implementation is provided on the GitHub Repository. A similar evaluation was done by Segal et al. in their paper using their PG-GAN results, so we opted not to redo the test for that model [8].

TABLE I
Classification Results

| S. No | Size and Type of Data | Accuracy on test set |
| --- | --- | --- |
| 1 | 1000 Real | 0.81 |
| 2 | 500 Real + 500 Synthesised | 0.73 |

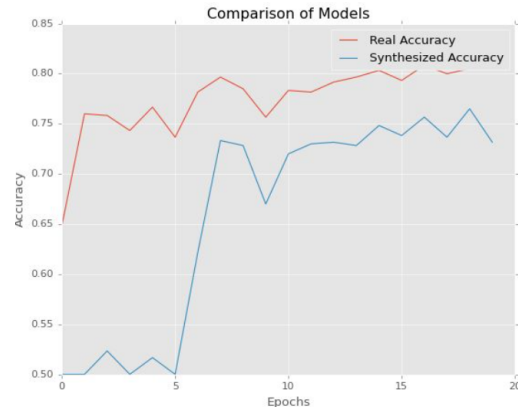

As the results show, the accuracy did fall when using synthesised images as part of the dataset, so the images generated are not quite on par with real images yet. However, it is encouraging to see that the model performed reasonably well despite half the data being synthesised.

There could be several reasons for the drop in accuracy, the most obvious one being that the images are simply not accurate enough to the real data. Another potential problem is overfitting. However, we have found little evidence of this happening, as the models were evaluated at several checkpoints and produced widely varying images.

Another possible reason for the observed discrepancy in the performance of the classification model on real versus synthesised images could be the limited number of images with diseases in the original dataset. Our generative model had many more No Finding images to train on than those with diseases, so perhaps it was not able to learn the features associated with each disease as well as it could have. The dataset has roughly 75% of its images with No Finding and only 25% for all the other classes combined.

## CONCLUSION

This experiment showed that although using Stable Diffusion for the purpose of generating synthetic medical data seems promising, there is still some work to be done. Perhaps the problems faced and be tackled in future work. The results were still promising, however, showing that even with synthetic data, classification models can perform reasonably well at CAD tasks. With the rapid improvement in image generation technology, there will no doubt soon come a time when using these methods will be able to increase the effectiveness of medical diagnosis through Deep Learning exponentially. (All the sample outputs, models etc, can be found on the official GitHub Repository)

## Acknowledgment

This paper is written as part of the final project of CS437 (Deep Learning) at the Lahore University of Management Sciences. It is an independent study with no sponsors.

## Author Information

**Muhammad Danyal Malik,** Undergraduate Student, Department of Computer Science, Lahore University of Management Sciences.

**Danish Humair,** Undergraduate Student, Department of Computer Science, Lahore University of Management Sciences.

## GitHub Repository

Link: https://github.com/mdanyalmalik/chest-xray-synthesis